\documentclass[traditabstract]{aa}
\usepackage{txfonts,natbib,graphicx}
\bibpunct{(}{)}{;}{a}{}{,}

\newcommand{\mxb}{MXB~1728--34}

\newcommand{\xmm}{{\it XMM-Newton}}
\def\ltsima{$\; \buildrel < \over \sim \;$}
\def\simlt{\lower.5ex\hbox{\ltsima}}
\def\gtsima{$\; \buildrel > \over \sim \;$}
\def\simgt{\lower.5ex\hbox{\gtsima}}

\begin{document}

\title{X-ray Spectroscopy of \mxb\ with \xmm}
\author{E.~Egron$^{1}$\thanks{mail to elise.egron@dsf.unica.it} \and 
T.~Di Salvo$^{2}$ \and L.~Burderi$^{1}$ \and A.~Papitto$^{1}$ \and 
L.~Barrag\'an$^{3}$ \and T.~Dauser$^{3}$ \and J.~Wilms$^{3}$ \and 
A.~D'A\`i$^{2}$ \and  A.~Riggio$^{1,4}$  \and R.~Iaria$^{2}$ \and N.~R.~Robba$^{2}$}
\institute{Dipartimento di Fisica, Universit\`a degli Studi di Cagliari, 
SP Monserrato-Sestu, KM 0.7, Monserrato, 09042 Italy
\and Dipartimento di Scienze Fisiche ed Astronomiche, Universit\`a di Palermo, 
via Archirafi 36, Palermo, 90123, Italy 
\and Dr. Karl Remeis-Sternwarte and Erlangen Centre for Astroparticle Physics,
Friedrich-Alexander-Universit\"at Erlangen-N\"urnberg, Sternwartstra$\ss$e 7, 
96049 Bamberg, Germany
\and INAF - Osservatorio Astronomico di Cagliari, Poggio dei Pini, Strada 54, 
09012 Capoterra (CA), Italy}

\abstract
{We have analysed an \xmm\ observation of the low mass X-ray binary and
atoll source \mxb. The source was in a low luminosity state during the 
\xmm\ observation, corresponding to a bolometric X-ray luminosity of 
$5 \times 10^{36}\ d_\mathrm{5.1\ kpc}^2\ \mbox{erg s$^{-1}$}$. 
The 1--11 keV X-ray spectrum of the source, obtained combining data 
from all the five instruments on-board \xmm, is well fitted by a Comptonized 
continuum. Evident residuals are present at $6-7$~keV which are ascribed 
to the presence of a broad iron emission line. This feature 
can be equally well fitted 
by a relativistically smeared line or by a 
self-consistent, relativistically smeared, reflection model. Under the
hypothesis that the iron line is produced by reflection from the inner 
accretion disk, we can infer important information on the physical parameters 
of the system, such as the inner disk radius, $R_\mathrm{in} = 25-100$~km, 
and the inclination of the system, $ 44^\circ < i < 60^\circ$.}

\titlerunning{XMM-Newton observation of MXB~1728--34}
\authorrunning{Egron et al.}

\maketitle

\section{Introduction}

Broad iron emission lines in the energy range 6.4$-$6.97 keV have been detected 
in high energy resolution spectra of many X-ray sources containing a compact 
object, such as active galactic nuclei \citep[e.g.,][]{Tanaka_95,Fabian_00}
and X-ray binary systems containing a stellar-mass black hole 
\citep[e.g.,][for a review]{Miller_02,Miller_07}, or a weakly-magnetized 
neutron star \citep[e.g.,][and references therein]{Bhattacharyya_07,
Cackett_08,disalvo_09,Papitto_09,Iaria_09,dai_09}. 
Identified with fluorescent K$\alpha$ transition of iron at different 
ionization states, these lines are generally interpreted in terms of reflection of the
central  hard X-ray emission on the accretion disk
\citep{Fabian_89}. 
Under this hypothesis, these lines are made broad and asymmetric by 
Doppler and relativistic effects induced by the Keplerian motion in the 
accretion disk near the compact object. The shape of the line 
is therefore an almost unique proxy of
the innermost accretion disk close to 
the compact object \citep[see][for a review]{Reynolds_03}, and, in 
particular, on the inner disk radius. It also indicates the inclination 
angle of the system and the ionization state of the reflecting matter. 
Other reflection components like absorption edges and the Compton
hump, usually observed between 20$-$40 keV, are also expected resulting 
from photoelectric absorption and Compton scattering of the main 
Comptonization continuum on the accretion disk matter.

\mxb\ (4U~1728$-$34, GX~354$-$0) is a low mass X-ray binary
containing a weakly magnetized accreting neutron star. The optical
counterpart of this ``galactic bulge'' source has not been identified
yet, due to the high optical extinction towards the Galactic center. 
Discovered in 1976 with the Small Astronomy Satellite SAS-3
\citep{Lewin_76,Hoffman_76}, this source belongs to the so-called atoll 
class \citep{Hasinger_89} and shows frequent type-I X-ray bursts 
\citep[e.g.,][]{Basinska_84} which are caused by thermonuclear flashes on 
the neutron star surface. 
Furthermore, double-peaked burst profiles 
have been observed \citep{Hoffman_76} which are explained as caused by 
photospheric radius expansion during the burst \citep{Taam_82}.
These bursts have been used to constrain the distance to the source, 
between 4.1 and 5.1 kpc \citep{disalvo_00,Galloway_03}. 
The power spectrum of \mxb\ displays kilohertz quasi-periodic
oscillations (QPOs) in the persistent emission, and a nearly
coherent oscillation at $\sim 363$ Hz during some bursts that has been 
interpreted as the spin frequency of the neutron star \citep{Strohmayer_96}.

Spectral analysis of \mxb\ has been performed in the past using data 
from different satellites, such as {\it Einstein} \citep{Grindlay_81}, 
{\it SAS-3} \citep{Basinska_84}, {\it EXOSAT} \citep{White_86}, {\it SIGMA}
\citep{Claret_94}, {\it ROSAT} \citep{Schulz_99}, and more recently using
{\it BeppoSAX} \citep{Piraino_00,disalvo_00}, {\it ASCA} \citep{Narita_01}, 
{\it RXTE}, {\it Chandra} \citep{dai_06}, {\it INTEGRAL} \citep{Falanga_06}, 
and {\it XMM-Newton}    \citep{Ng_10}.
The X-ray spectrum is generally composed of a soft and a hard component.
The first one can be described by a blackbody or a multicolor disk blackbody, 
and may originate from the accretion disk. 
The second one can be fitted either by a Comptonized spectrum or a thermal 
Bremsstrahlung. The Comptonized model seems to be more realistic, since it 
is efficiently produced by soft photons coming from the neutron star
surface and/or boundary layer between the accretion disk and the neutron 
star and/or the accretion disk, which are up-scattered by electrons in 
a hot corona.
                                            
A broad emission line at 6.7 keV has been often detected in the X-ray 
spectra of this source and has been interpreted as emission from highly 
ionized iron \citep[e.g.,][]{disalvo_00}. The large width of the line 
suggests that it could come from an ionized inner accretion disk 
\citep{Piraino_00}, or alternatively it could be emitted from a strongly 
ionized corona. \citet{dai_06} have proposed an alternative model to
describe the iron line region using two absorption edges associated with
ionized iron instead of a Gaussian line.

In this paper, we present a spectral analysis of high energy resolution 
data taken by \xmm\ on 2002 October 3rd using all the five X-ray instruments
on-board this satellite. The EPIC-pn data have already been published 
by \citet{Ng_10}, in a 'catalog' paper dedicated to the study of the iron 
line in 16 neutron star LMXBs observed by \xmm.\ 
Here we present  a different approach to the analysis of these data, 
since for the first time we fitted \xmm\ data of MXB 1728-34 with a
self-consistent modelling of the continuum emission and of the reflection 
component.
We also tried several models to fit the iron line profile, such as 
diskline or relline, which are different from the models proposed by 
\citet{Ng_10} (Gaussian or Laor). All our results favor the 
relativistic nature of the line profile.
We fitted simultaneously the spectra from all the five X-ray instruments 
onboard \xmm, while just the pn data are analysed by \citet{Ng_10}.

\section{Observation and data reduction}

\mxb\ was observed by \xmm\ on 2002 October 3rd for a total on-source 
observing time of 28 ks. The observation details for the instruments 
on-board \xmm\, including the European Photon Imaging Camera 
\citep[EPIC-pn,][]{struder_01}, the MOS1 and MOS2 cameras 
\citep[][]{turner_01}, and the Reflection Grating Spectrometer 
\citep[RGS1 and RGS2,][]{denherder_01} are presented in the Table~1. 
The Optical Monitor \citep[OM,][]{mason_01} was not active during this 
observation.

\begin{table}
\begin{minipage}[t]{\columnwidth}
\caption{Instrument modes, filters and exposure times.}
\label{tabps}
\centering
\renewcommand{\footnoterule}{}  
\begin{tabular}{llcc}
\hline \hline
   Instrument & Mode & Filter & Exposure (ks) \\
   \hline
   pn & Timing mode & Thick & 26.9 \\
   MOS1 & Timing mode & Thick & 27.5 \\
   MOS2 & Timing mode & Thick & 27.5 \\
   RGS1 & Standard spectroscopy & - & 28.1 \\
   RGS2 & Standard spectroscopy & - & 28.1 \\
   \hline
\end{tabular}
\end{minipage}
\end{table}

The $1.5-12$~keV lightcurve of the All Sky Monitor 
(ASM) on board {\it RXTE} extracted within $\sim 450$ days the \xmm\ observation indicates that the 
source was not in a high activity state, since it shows an average 
count rate of about 3 counts/s. 

\begin{figure}[t]
\includegraphics[width=9.2cm]{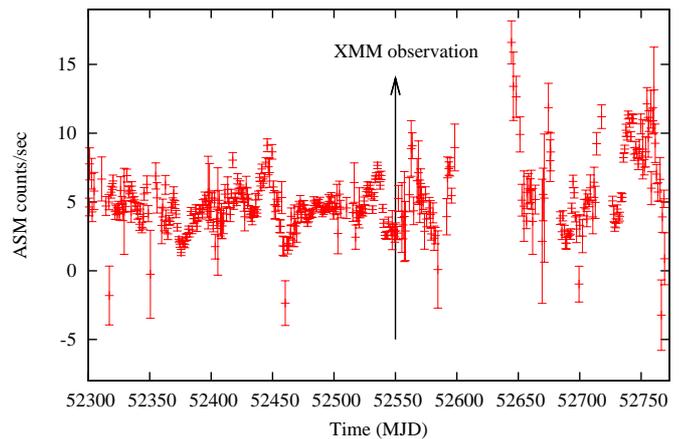}
\caption{{\it RXTE}/ASM lightcurve covering the 1.5$-$12 keV energy range. 
The vertical line indicates the time of the \xmm\ observation performed
on 2002 October 3rd.}
\label{fig:lc_ASM}
\end{figure}


The \xmm\ data were processed using the Science Analysis Software
v.~9.
The EPIC-pn camera was operated in timing mode to prevent photon
pile-up. We created a calibrated photon event file using the pn
processing tool \textsc{epproc}. Before extracting the spectra, we
checked for contamination from background solar flares by producing a
lightcurve in the energy range 10$-$12 keV. There were no 
high background periods during this observation. 
We used the task \textsc{epfast} to correct rate-dependent CTI effects 
in the event list. The source spectrum was extracted from a rectangular 
area, covering all the pixels in the Y direction, and centered on the 
brightest RAWX column (RAWX=38), with a width of 13 pixels around the 
source position (because 90\% of the source counts up to 9 keV is encircled 
by 53 arcsec, and 1 pn pixel is equivalent to 4.1 arcsec). We selected 
only events with $\mathrm{PATTERN} \leq 4$ (single and double pixel events) 
and FLAG=0 as a standard procedure to eliminate spurious events.  
We extracted the background away from the source (in the $\mathrm{RAWX}=6-18$).
We also checked that pile-up did not affect the pn spectrum using the 
task \textsc{epatplot}. The total count rate registered by EPIC-pn CCDs 
was around 110 count/s, and was 64 count/s in the 2.4$-$11 keV range,
slightly increasing (by a 5\%) during the observation.

The MOS data were also taken in timing mode and processed with the 
routine \textsc{emproc} to produce calibrated event list files. 
The source spectra were extracted from a rectangular box centered on
RAWX=320 (MOS1), and on RAWX=308 (MOS2), selecting an area 30 pixels 
wide around the source position, and covering 722 pixels on
the Y (PHA) direction. Only events corresponding to $\mathrm{PATTERN} 
\leq 12$ and FLAG$ = 0$ were selected, corresponding to standard filters. 
The background spectra were extracted far from the source, centered 
in the column RAWX$=240$. We checked that the MOS spectra were not 
affected by pile-up. The count rates were estimated around 30 counts/s 
for each MOS unit (20 counts/s considering 2.4$-$11 keV energy range). 


Spectral channels of EPIC-pn and MOS spectra have been rebinned to have  
3 channels per energy resolution element and at least 25 counts per 
energy channel.

The two RGS were operated in the standard spectroscopy mode. RGS data
were processed using the \textsc{rgsproc} pipeline to produce
calibrated event list files, spectra and response matrices. The count 
rates measured by RGS1 and RGS2 were around 2.5 and 3.5 counts/s,
respectively. The RGS data were rebinned in order to have a minimum of 25
counts per energy channel.

\begin{table}
\begin{minipage}[t]{\columnwidth}
\caption{Best fitting parameters of the continuum emission for 
the \xmm\ pn, MOS1, MOS2, RGS1 and RGS2 spectra of \mxb.}
\label{tabps}
\centering
\renewcommand{\footnoterule}{}  
\begin{tabular}{lr}
\hline \hline
   \textbf{Parameter} & \textbf{Value} \\
   \hline
   $N_\mathrm{H} ~ (\times 10^{22}cm^{-2})$ & $ 2.4 \pm 0.1 $ \\
   $kT_\mathrm{seed}$ (keV) & $ 0.59 \pm 0.02 $ \\
   $kT_\mathrm{e}$ (keV) & $ 2.74 \pm 0.04 $ \\
   $\tau$ & $ 16.5 \pm 0.2 $ \\
   Norm & $ 9.54 \pm 0.02 $ \\
   Flux 2.0$-$10.0 keV (pn) & 8.06 \\
   Flux 2.0$-$10.0 keV (MOS) & 8.17 (MOS1) - 8.07 (MOS2) \\
   Flux 1.0$-$2.0 keV (RGS) & 0.186 (RGS1) - 0.185 (RGS2) \\
   Total $ \chi^{2}$ (d.o.f.) & 1732 (903) \\
   \hline
\end{tabular}
\end{minipage}
\textit{Note: The model used to fit the continuum is 
\textbf{cons*phabs*compTT}. The absorbed flux is in units of 
$ 10^{-10} erg \,cm^{-2} \,s^{-1}$.}
\end{table}

\section{Spectral analysis}

Data were fitted by using \textsc{Xspec} \citep{arnaud_96} v.12.5.1. 
All uncertainties are given at the 90\% confidence level 
($\Delta \chi^{2} = 2.706$).
We fitted simultaneously the broad band energy spectra of the source 
obtained from all the five instruments. 
Considering the best calibration ranges of the different 
detectors, the data analysis from EPIC-pn, MOS1 and MOS2 cameras 
was restricted to the energy range 2.4$-$11 keV. This excluded the region
around the detector Si K-edge (1.8 keV) and the mirror Au M-edge (2.3 keV)
that could affect our analysis. This problem was already noticed for the EPIC-pn 
observations performed in timing mode, e.g., \citet{boirin_05}, \citet{Iaria_09}, 
\citet{dai_09}, \citet{Papitto_09}, \citet{dai_10}, \citet{papitto_10}.
We used only the RGS1 and RGS2 data 
between 1$-$2 keV to constrain the softest band.

The different cross calibrations of the five instruments 
were taken into account by including normalizing factors in the model. 
These factors were fixed to 1 for pn and kept free for the other 
instruments.

We first fit the continuum with a thermal Comptonized model
using \textsc{compTT} \citep{titarchuk_94}, modified at low
energy by the interstellar photoelectric absorption modelled by
\textsc{phabs} using photoelectric cross-sections of \citet{balucinska_98}
with a new He cross-section based on \citet{yan_98} 
and standard abundances of \citet{anders_89}. The $ \chi^{2}$/degrees of 
freedom (d.o.f.) of 
the fit was large, 1732/903\label{fig:cont}.
We then tried to add a blackbody component (\textsc{bbody} model) 
to improve the fit. 
The addition of this component resulted to be not statistically 
significant, thus we decided not to include the blackbody in our model. 
The values of the parameters of the continuum emission are reported 
in Table 2.

With respect to this continuum model, an excess was present in the 
residuals between 5.5 and 8 keV, probably indicating the presence of 
iron discrete features. 
The fit was improved by adding a broad iron emission line, modelled by 
a simple Gaussian line (Model 1 in Table 3), centered at 6.6 keV with the 
$\sigma$ parameter frozen at 0.6 keV. Indeed, the width of the line was poorly 
constrained, however when we fitted the MOS data independently from the pn data, 
a significantly broadened line was detected, with a lower limit on $\sigma$ of 
0.55 keV. With the addition of the Gaussian, the fit gave a 
$ \chi^{2}$/d.o.f. = 1489/901 (resulting to a significant improvement of the fit, with a
$\Delta \chi^2 = 243$ for the addition of two parameters). 
We also tried to fit the line with a combination of two emission lines instead of a broad line, 
but the $ \chi^{2}$ was worse ($ \chi^{2}$/d.o.f. = 1531/901). So this broad line does 
not result from a blending of iron line at different ionization state.

\begin{table*}[t]
\caption{Results of the fitting of \mxb\ \xmm\ data with different models. 
The five models reported in this table mainly differ for the modelling 
of the iron feature. 
In the first four models the X-ray continuum is fitted with 
\textbf{const*phabs*CompTT}, to which a Gaussian line (Model 1), 
or a diskline (Model 2), or a relativistic line (Model 3) or two edges 
(Model 4) are used to fit residuals in the iron K-shell range. 
Model 5 includes a different Comptonization continuum model and 
a self-consistent, relativistically smeared, reflection component: 
\textbf{const*phabs*(nthComp + rdblur * reflionx)}.}
\label{tabps}
\centering
\renewcommand{\footnoterule}{}  
\begin{tabular}{llccccc} 
\hline \hline

& & \textbf{Model 1:} & \textbf{Model 2:} & \textbf{Model 3:} & \textbf{Model 4:} & \textbf{Model 5:} \\

\textbf{Component}& \textbf{Parameter} & \textbf{Gaussian} & \textbf{Diskline} & 
\textbf{Relline} & \textbf{Two edges} & \textbf{Reflection}\\

   \hline
   edge & E edge (keV) & - & - & - & $ 7.5 \pm 0.1 $ & - \\
   edge & $\tau ~ ( \times 10^{-2})$ & - & - & - & $ 6\pm 1 $ & - \\
   edge & E edge (keV) & - & - & - & $8.49^{+0.09}_{-0.07}$ & - \\
   edge & $\tau ~ ( \times 10^{-2})$ & - & - & - & $ 6 \pm 1 $ & - \\
   phabs & $ N_\mathrm{H} ~ (\times 10^{22}cm^{-2})$ & $ 2.2 \pm 0.1 $ & $ 2.2 \pm 0.1 $ 
 & $ 2.2 \pm 0.1 $ & $ 2.4 \pm 0.1 $ & $ 2.7 \pm 0.1 $ \\
   compTT & $ kT_\mathrm{seed} $ (keV) & $ 0.69 \pm 0.02 $ & $0.69^{+0.01}_{-0.02}$ 
& $ 0.68 \pm 0.02 $ & $ 0.62 \pm 0.02 $ & - \\
   compTT & $ kT_\mathrm{e}$ (keV) & $ 3.2 \pm 0.1 $ & $ 3.3 \pm 0.1 $ 
& $ 3.3 \pm 0.1 $ & $ 3.1 \pm 0.1 $ & - \\
   compTT & $ \tau $ & $14.1^{+0.4}_{-0.5}$ & $13.9^{+0.5}_{-0.3}$ 
& $ 14.0^{+0.4}_{-0.2} $ & $ 15.2  \pm 0.3 $ & - \\
   compTT & Norm $ (\times 10^{-2})$ & $ 7.6 \pm 0.3 $ & $ 7.4 \pm 0.3 $ 
& $ 7.5^{+0.3}_{-0.4}$ & $ 8.3 \pm 0.3 $ & - \\
   nthComp &$ \Gamma $ & - & - & - & - & $ 1.84^{+0.04}_{-0.01} $ \\
   nthComp & $ kT_\mathrm{e}$ (keV) & - & - & - & - & $ 4.9^{+1.4}_{-0.7} $ \\
   nthComp & $ kT_\mathrm{bb}$ (keV) & - & - & - & - & $ 0.71^{+0.03}_{-0.01} $ \\
   nthComp & Norm $ (\times 10^{-2})$ & - & - & - & - & $ 4.9 \pm 0.2$ \\
   Gauss & E (keV) & $ 6.57 \pm 0.05 $ & - & - & - & -\\
   Gauss & $ \sigma $ (keV) & 0.6 (frozen) & - & - & - & -\\
   Gauss & Norm $ (\times 10^{-4})$ & $ 8.8 \pm 1 $ & - & - & - & -\\
   diskline & E (keV) & - & $6.45^{+0.05}_{-0.07}$ & - & - & -\\
   diskline & Betor & - & $(-2.8)^{+0.2}_{-0.3}$ & -& - & - \\
   diskline & $ R_\mathrm{in}$ (GM/$ c^{2}$) & - & $18^{+3}_{-6}$ & - & - & -\\
   diskline & $ R_\mathrm{out}$ (GM/$ c^{2}$) & - & 1000 (frozen) & - & - & -\\
   diskline & i ($^{\circ}$) & - & 60 (frozen) & - & - & - \\
   diskline & Norm $ (\times 10^{-4})$ & - & $ 9.6 \pm 1 $ & - & - & - \\
   relline & E (keV) & - & -  & $ 6.43 \pm 0.07 $ & - & - \\
   relline & Index 1 & - & - & $ 2.8^{+0.2}_{-0.1}$ & - & - \\
   relline & i ($^{\circ}$) & - & - & 60 (frozen) & - & - \\
   relline &  $ R_\mathrm{in}$ (GM/$ c^{2}$) & - & - & $ 19^{+3}_{-4}$ & - & - \\
   relline & $ R_\mathrm{out}$ (GM/$ c^{2}$)  & - & - & 1000 (frozen) & - & - \\
   rdblur & Betor  & - & - & - & - & -2.8 (frozen) \\
   rdblur & $ R_\mathrm{in}$ (GM/$ c^{2}$)  & - & - & - & - & $20^{+29}_{-6}$ \\
   rdblur & $ R_\mathrm{out}$ (GM/$ c^{2}$)  & - & - & - & - &  1000 (frozen)\\
   rdblur & i ($^{\circ}$) & - & - & - & - & $> 44$ \\
   reflion & Fe/Solar & - & - & - & - & 1 (frozen)\\
   reflion & $ \Gamma $ & - & - & - & - & $ 1.84^{+0.04}_{-0.01} $\\
   reflion & $ \xi $ & - & - & - & - & 660 (frozen) \\
   reflion &  Norm $ (\times 10^{-6})$  & - & - & - & - & $ 2.2^{+0.5}_{-0.4}  $\\
   & Total $ \chi^{2} $ (d.o.f.) & 1489 (901) & 1463 (899) 
& 1464 (899) & 1519 (899) & 1463 (900) \\
   \hline
 
\end{tabular}
\vskip 0.5cm
\end{table*}

We then tried to substitute the Gaussian at 6.6 keV with a diskline profile 
and obtained a slightly better result with a $\chi^{2}$/d.o.f. = 1463/899 
\label{fig:diskl}. The data and the residuals obtained using this model are
shown in Fig. 2. Due to the large uncertainties on the outer radius of the 
disk and on the inclination of the system if they were let free, we have 
frozen the first one at 1000 $R_\mathrm{g}$ and the inclination at $60^\circ$ 
(the source does not show any dip in its lightcurve, implying $i < 60^\circ$).
The improvement of the fit corresponds to $ \Delta \chi^{2} = 26 $ for the 
addition of two parameters (the F-test gives a probability of chance 
improvement of about $10^{-4}$). 

We think that the F-test gives a reliable result in this case.
In fact, according to \citet{protassov_02}, the conditions that have 
to be satisfied in order to properly use the F-test are: i) the two 
models that are being compared must be nested; ii) the null values of 
the additional parameters should not be on the boundary of the set of 
possible parameter values.
We think that a comparison between a Gaussian and a diskline satisfies 
these conditions, since a Gaussian profile can be obtained by a diskline 
(which is indeed given by a narrow, $\sigma=0$, Gaussian multiplied 
by the kernel \textsc{rdblur}) and a good approximation of a Gaussian 
profile is obtained by a diskline for values of the parameters not
at their boundary.  
However, in order to assess the significance of the relativistic line 
smearing on a statistical basis, we used an other statistical method 
based on the Posterior Predictive p Values described by 
\citet{hurkett_08}. For simplicity, we have restricted our data to 
two instruments (pn and MOS2). These data were fitted  with Model 1 
and Model 2, respectively. We obtained an improvement of  the fit
corresponding to a $\Delta \chi^{2}$ of 18 for the addition of 2 
parameters when we substituted Model 1 with Model 2. Then, we simulated 
200 pn and MOS2 spectra according to Model 1, which were fitted in a 
second step using Model 2. The $\Delta \chi^{2}$ was registered for 
each simulation. Among the 200 simulations, we found 2 times a 
$\Delta \chi^{2}$ higher than 18. So the probability of chance 
improvement we get from these simulations is 1\%, which is 
in agreement with the 0.75\% calculated by the F-test performed on 
the real data restricted to two instruments. Therefore we can 
conclude that the diskline model is preferred since it gives a 
probability of chance improvement of the fit of about $10^{-4}$ 
using all the instruments. 

This model (Model 2 in Table 3) gives an 
estimate of the inner radius of the disk $R_\mathrm{in}\sim 18$ 
$R_\mathrm{g}$ ($R_\mathrm{g}$ = GM/$c^{2}$ is the Gravitational radius). 
Fixing the inclination to lower values (i.e. $i < 60^\circ$) , worse $\chi^{2}$ were 
obtained with the other diskline parameters drifting towards lower 
$R_\mathrm{in}$, higher rest-frame energies, and higher, in absolute value, 
emissivity indexes. 
We searched for an absorption edge in the energy range 7$-$10 keV, but
none was significantly detected.

To fit the iron line, we also used a new model for a relativistically 
distorted disk line, called \textsc{relline}\footnote[1]{http://www.sternwarte.uni-erlangen.de/research/relline/} \citep{dauser_10}, which calculates line profiles taking into account all the relativistic 
distortions in a disk around the compact object. 
We also fixed the outer radius of the disk and the inclination of the 
system to the same values used in the diskline model. The $\chi^{2}$/d.o.f. 
obtained in these conditions is 1464/899. The best fit line parameters
obtained in this way are perfectly consistent with those obtained using the
diskline model. In particular this model (Model 3 in Table 3) estimates 
the inner radius of the disk to be $R_\mathrm{in}\sim 19 $ $R_\mathrm{g}$. 

We tried an alternative model for the iron features \citep[see][]{dai_06} 
using two absorption edges (instead of an emission line) which 
are found at 7.50 keV ($\tau \sim 0.06$) and 8.49 keV ($\tau \sim 0.06$),
associated to mildly and highly ionized iron, respectively. 
The $\chi^{2}$/d.o.f. for this fit is 1519/899 (which has to be compared to 
1489/901 that we obtained fitting the iron line with a Gaussian (Model 1) 
or to 1463/899 that we obtained fitting the iron line with a diskline 
(Model 2)). Therefore this model (called Model 4 in Table 3) gives a worse
fit of the iron features than the previous ones.

In order to test the consistency of the broad iron line with
a reflection component, we fitted the data using \textsc{reflion},
a self-consistent reflection model including both the reflection continuum 
and the corresponding discrete features \citep{Ross_05}, in 
addition to a thermally comptonized continuum modelled with \textsc{nthComp}, 
by \citet{Zdziarski_96}, and extended by \citet{Zycki_99}), 
instead of \textsc{compTT}. The $\chi^{2}$/d.o.f. was 1555/901, without
the inclusion of relativistic smearing. 
Since the iron line was found to be significantly broad in the previous 
models (Model 1, 2, and 3), we added the relativistic smearing using 
the \textsc{rdblur} component. 
The addition of this component to the model constitutes Model 5 in Table 3 
and led to a $\chi^{2}$/d.o.f. = 1466/899. The decrease of the $\chi^2$ 
for the addition of the relativistic smearing was $\Delta \chi^2 = 89$ 
for the addition of 2 parameters (corresponding to an F-test probability of
chance improvement of $\sim 10^{-12}$).
Assuming that iron has a solar abundance and freezing the emissivity
betor index to -2.8 (value obtained with the diskline model) and the ionization parameter 
$ \xi = L_\mathrm{X} / (n r^{2})$ to 660 (this parameter tends to take high values but is unstable 
during the fit, that is why we preferred to freeze it), where $ L_\mathrm{X} $ is the 
ionizing X-ray luminosity, \textit{n} is the electron density in the reflector, 
and \textit{r} the distance of the reflector to the emitting central source,
the inner radius has been estimated again to be 20 $R_\mathrm{g}$ and a 
lower limit to the inclination angle was found to be $44^\circ $.
In Fig. 3 we plot the residuals obtained with this self-consistent reflection model, in comparison with the continuum
and the \textsc{relline} model using the EPIC-pn data.

The residuals found with respect to the different models described above do not show any evident systematic trend; the large
$\chi^{2}$ could be due to mismatches in the cross-calibration between the different instruments (for more details, see the cross-calibration document available on the \xmm\ webpage\footnote[2]{http://xmm2.esac.esa.int/docs/documents/CAL-TN-0052.ps.gz}) or to unresolved and unfitted features.




\begin{figure}[t]
\includegraphics[width=9.3cm]{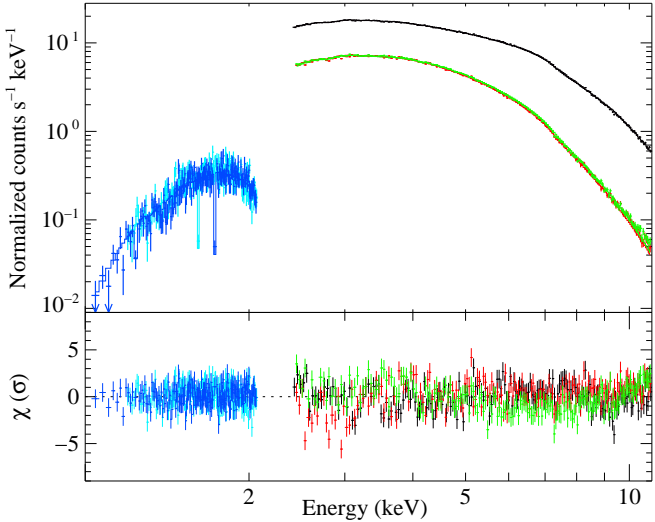}
\caption{Top panel: EPIC-pn (black), MOS1 (red), MOS2 (green), RGS1 (cyan), 
RGS2 (blue) data points of \mxb\ in the range 1$-$11 keV. 
Bottom panel: Residuals (Data-Model) in unit of sigmas for the diskline model (Model 2 in Table 3).}
\label{fig:5mod}
\end{figure}

\begin{figure}[t]
\includegraphics[width=9.3cm]{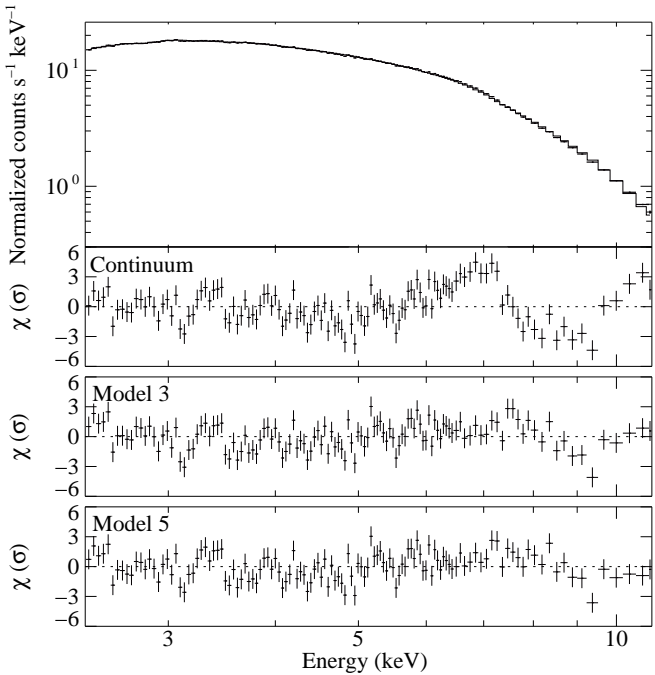}
\caption{Top panel: EPIC-pn data points of \mxb\ in the range 2.4$-$11 keV. 
Bottom panels: Residuals (Data-Model) in unit of sigmas for the continuum 
model reported in Table 2, for Model 3 including a relativistic line 
(\textsc{relline}), and for Model 5 using a relativistic reflection component 
(\textsc{reflion}), respectively. 
Data have been rebinned for graphical purposes.}
\label{fig:rel_ref}
\end{figure}


\section{Discussion}

We performed a spectral analysis of \mxb\ observed by \xmm\ on 2002 
October 3rd in the 1$-$11 keV energy range. The best fit continuum 
model consists of an absorbed Comptonized component; the addition of 
a soft blackbody component does not improve the fit significantly. 
With respect to this continuum model, evident residuals are present 
at $6-8$ keV, which can be fitted 
either by a relativistic line (such as diskline or \textsc{relline}) or by a 
self-consistent relativistic reflection model.

\subsection{The continuum emission}

The X-ray spectra of black-hole and neutron star in X-ray binaries 
are generally described by models that include a soft/thermal
and a hard/Comptonized component; the electron temperature of the Comptonized
component significantly 
decreases when the source transits from the hard to the soft 
state, while its optical depth increases. For hard spectra, Comptonization is unsaturated and the spectrum may
be approximated by a cutoff power-law. 
In soft states, Comptonization is saturated and the spectral shape can be approximated 
with a blackbody emission at the electron temperature.
During the observation with \xmm,\ \mxb\ was in a low luminosity state, 
and presumably an unsaturated Comptonization is the expected spectral shape. 
We chose a continuum which consists of an absorbed comptonization model 
(\textsc{comptt} or \textsc{nthComp}) because these models gave the best fit 
to the broad-band ($0.1-200$ keV) {\it BeppoSAX} spectrum of \mxb\ 
\citep{Piraino_00,disalvo_00}.

A different continuum was used by \citet{Ng_10} to fit the 
sample of neutron star LMXBs. Most of the observations, including \mxb,\ 
were well fitted by a blackbody plus a disk blackbody component, absorbed 
by neutral interstellar matter. The temperature of the blackbody 
and of the disk blackbody component had values between 1.5 and 2.8 keV 
and between 0.6 and 1.3 keV, respectively, except for \mxb\ for which 
temperatures were found between $3.8^{+8.4}_{-1.1}$ keV and $1.9 \pm 0.3 $
keV, respectively. 
An alternative model consisting of a blackbody and a power-law was also used 
to fit the continuum. However, in this case the authors warn the reader 
about the fact that different continuum parameters were found when fitting 
the residuals at the Fe band with different models, meaning that line fit 
may not be realistic. Moreover, the authors note that \mxb\ was in a low
luminosity state during the \xmm\ observation and that for such dim sources 
Compton scattering is expected to play an important role but was not included 
in their models.
In our case, fitting the X-ray continuum with a Comptonization component, the parameters of the 
continuum do not change significantly when we add a gaussian or a 
relativistic line (such as diskline or \textsc{relline}) to the 
continuum. 
Our choice of the continuum modelling allows us to better constrain the profile of the
broad iron line. In this context, the choice of a particular fitting to the
line does not result in sensible changes of the parameters determining the
continuum emission.
Moreover when we tried  a self-consistent reflection model, we found again 
similar values of the spectral parameters, which again indicates that the 
fit is stable and does not depend on the particular choice of the model 
used to fit the iron line.

The equivalent hydrogen column inferred from the Galactic photoelectric 
absorption component, $N_\mathrm{H} \sim 2.8 \times 10^{22}$~cm$^{-2}$, is 
in agreement with typical values for this source ($N_\mathrm{H} \sim 2.6-2.7 
\times 10^{22}$~cm$^{-2}$, \citep{disalvo_00,Piraino_00,dai_06}). 
The Comptonized component can be produced by inverse Compton scattering 
from relatively hot electrons ($kT_\mathrm{e} \sim 3.3$~keV using the 
\textsc{compTT} model or $kT_\mathrm{e} \sim 4.8$~keV using the 
\textsc{nthComp} model) off soft photons ($kT_\mathrm{seed} \sim 0.7$~keV 
using the \textsc{compTT} model or the \textsc{nthComp} model).
The seed photons for Comptonization are compatible with coming from the
neutron star surface. In fact we can estimate the size of the emitting 
region of the soft photons, using the formula given by \citet{intzand_99}. 
For this, one assumes that the bolometric luminosity of the soft 
photons is equal to the corresponding blackbody luminosity at the Wien 
temperature. 
The relative gain $y = 5.01$ (for a spherical geometry) takes into account 
the energy gained by the photons scattered off relativistic electrons through 
the inverse Comptonization process. This leads to a value $R_\mathrm{seed} = 
4.9 d_\mathrm{5.1}$~km, where $d_\mathrm{5.1}$ is the distance in units of
5.1 kpc, considering the unabsorbed flux that we extrapolated in 
the energy range 0.1--150 keV, $F_\mathrm{bol} \sim 1.5(2) \times 10^{-9}$ 
ergs cm$^{-2}$ s$^{-1}$.

Usually a soft blackbody component is required to fit the broad band 
X-ray spectra of LMXBs, most frequently interpreted as emitted by an 
accretion disc. This component is not significantly detected in the 
\xmm\ spectrum. 
This may be ascribed to the relatively low X-ray luminosity of the source
during the \xmm\ observation, specially in the soft band. 
Also the bolometric X-ray flux in the range 0.1$-$150 keV 
implies a bolometric X-ray luminosity of $L_\mathrm{X} \sim 5 \times 10^{36} 
d^{2}_{5.1}$ erg s$^{-1}$,
corresponding to 2\% of the Eddington luminosity, that is 
$L_\mathrm{Edd} = 2.5 \times 10^{38}$ erg s$^{-1}$ for a 1.4 
$\mathrm{M_{\odot}}$ neutron star \citep[e.g.,][]{vanparadijs_94}. 
In fact, during the high/soft state the disk is expected
to be very close to the compact object, while in the low/hard state 
the disc should be truncated far from the compact object, and therefore  
its contribution is expected to be less important. 
We therefore conclude that the blackbody component is just too weak
to be detected. This is in agreement with the results obtained from
the fit of the iron feature with a reflection model. This indicates
that the inner accretion disk is probably truncated far from the 
neutron star ($R_\mathrm{in} > 25$~km), and with a relatively high value
for the system inclination with respect to the line of sight estimated 
at $ 44^\circ < i < 60^\circ $ which would further reduce the disk
luminosity with respect to the Comptonized component in the hypothesis 
that the last one has a spherical geometry around the compact object.

\subsection{The iron line emission}

Recently \citet{Ng_10} and \citet{cackett_10} presented a spectral
analysis of a sample of neutron star LMXBs observed by {\it Suzaku} and 
\xmm,\ respectively, with particular interest on the iron discrete features 
in these sources. While \citet{cackett_10} conclude that Fe K line profiles 
are well fitted by a relativistic line model for a Schwarzschild metric in 
most cases and imply a narrow range of inner disk radii ($6-15$ $GM/c^{2}$),  
\citet{Ng_10}  conclude there is no evidence for asymmetric (relativistic) 
line profiles in the \xmm\  data, although the line profiles (fitted with 
a simple Gaussian or a laor model) again appear quite broad, with Gaussian 
sigmas ranging between 0.17 up to 1.15 keV.
Another aspect relevant in this context is 
the impact of photon pile-up on relativistic disk lines and continuum spectra.
As shown in \citet{Miller_10}, while severe photon pile-up distorts relativistic disk lines
and the disk continuum is opposing ways, the line shape does not sensibly depend on the
pile-up fraction when its effect is modest.

We tried different models to fit the iron line profile. 
Although a Gaussian line provides an acceptable fit of the line profile, we tried
to physically interpret its large width using models for a relativistically smeared 
line in an accretion disk.

Using the diskline profile (Model 2) we find the line centroid energy 
at 6.45 keV, compatible with a fluorescent $K\alpha$ transition from 
mildly ionized iron (\ion{Fe}{I$-$XX}). The inner radius is in the 
range 12$-$21~$R_\mathrm{g}$. 
The line profile appears therefore to be significantly broad and compatible
with a diskline profile. 
The results obtained by using a relativistic line profile corresponding 
to Model 3, which uses the more recent  \textsc{relline}, instead of 
diskline, are perfectly consistent with the diskline model. The inner 
radius is estimated to be in the range 15$-$22~$R_\mathrm{g}$.

The line profile can be equally well fitted using a self-consistent 
relativistic reflection model (Model 5). 
The addition of the \textsc{rdblur} component improves significantly the $\chi^{2}$.
This indicates that the line is indeed broad, and that the width of the 
line is in agreement with a relativistic smearing in the disk.
The value of the inner radius 
is again consistent with that found using a diskline or a relline profile, 
even if the uncertainty is larger 
($R_\mathrm{in} = 13$-$43$~$R_\mathrm{g}$). For a neutron star 
mass of $1.4~\mathrm{M_\odot}$, the inner disk radius is in the range 
25$-$100~km from the neutron star center, and so the disk would 
be truncated quite far from the neutron star surface. The inclination angle
of the system with respect to the line of sight is found to be $> 44^\circ$, 
which is still compatible with the absence of dips
in its lightcurve (which implies $i < 60^\circ$).

We also attempted to fit the iron feature using an alternative model 
(Model 4), consisting of two absorption edges instead of a broad emission 
line \citep[see][]{dai_06}. Their energies correspond to moderately 
ionized iron, Fe from IX to XVI, and highly ionized iron, Fe XXIII, 
respectively. 
The fit was not so good as with the previous 
models, since the corresponding $\chi^2$ was larger by 56 with the same
number of degrees of freedom. 
We can compare these results with those obtained 
by \citep{dai_06}, using simultaneous {\it Chandra} and {\it RXTE} 
observations. The edges were found at slightly different energies of 
7.1 and 9.0 keV, corresponding to 
weakly (\ion{Fe}{I$-$V}) and highly ionized iron (\ion{Fe}{XXV$-$XXVI}), 
respectively. 
Although the fitting with two iron edges cannot be completely excluded yet, we think 
that the most probable explanation is that a couple of edges may mimic, 
given the relatively low statistics, the shape of a broad iron line. 
We therefore favor the interpretation of the iron feature as 
a broad and relativistic emission line produced in the accretion disk, 
since this gives a better fit of the \xmm\ spectrum and very 
reasonable values of the reflection parameters.

   


	
\begin{acknowledgements}
This work was supported by the Initial Training Network ITN 215212: Black 
Hole Universe funded by the European Community. \textbf {We thank the 
referee for useful comments which helped in improving the 
manuscript.}
\end{acknowledgements}

\bibliographystyle{aa}
\bibliography{biblio_1728}

\end{document}